# Lindeman's criterion: diamond graphitization temperature and its dependence on external pressure

V.N. Strekalov

Moscow State University of Technology "STANKIN", Vadkovsky per., 3a, Moscow 101472, Russia,

stvn@stankin.ru

*A simple model for the "diamond graphitization" type of phase transition is proposed, leading to a good agreement with the experimental data. The phase transition temperature dependence on the external pressure is found. A new phenomenon in diamond is predicted, which is a decrease of the temperature of graphitization with increase of external pressure.*

It has been often mentioned that diamond is a unique metastable crystal that has exceptional physical properties. Many of the diamond parameters have been studied albeit with occasional discrepancies and even contradictions between various sources [1–5]. In many cases the discrepancies are due to the fundamental distinctions of the samples.

It is well known that diamond can spontaneously [1, 6–8] transform into other forms of graphite (e.g. when the natural diamonds are faceted, small quantities of dark graphite-like material are observed [1, 6]).

At the present time, many experimental investigations have been reported carried out under quasi-equilibrium as well as under non-equilibrium conditions, in the presence of chemical agents or external influence (such as caused by electron, ion or laser beams, etc). However the theoretical description of the diamond transformations is far from complete.

It is usually pointed out in the theoretical works that only the initial stage of graphitization is considered, when only a thin surface layer of graphite is created. This limitation arises from the requirement of free volume, which is available near the surface or near lattice defects. Study of graphitization at longer times and in anisotropic systems requires for description of the process nonequilibrium transport equations (which is not accomplished yet) and requires additional approximations reducing, rather than improving, the result's reliability. Mainly, these are the mathematical difficulties and complexity of the graphitization process that forces the researchers to reduce their analysis to the isotropic models. The isotropic approximation is a standard, widely accepted model in such problems [3, 5, 8 - 10].

Several mechanisms of the "true" graphitization are discussed, see the review in [1].

The first, "high-temperature" graphitization mechanism assumes that the carbon atoms experiencing the thermal fluctuations, acquire the energy exceeding the energy of their surface binding. Such atoms leave the surface and then re-crystallize but now in a different allotropic form [3, 12, 13]. In this case the fluctuation has to exceed the energy of sublimation, variable from $\varepsilon_1 =$ 7.6 eV/atom up to $\varepsilon_1 = 10.9$ eV/atom (for different directions) [3].

The second true, "low-temperature" mechanism is evident from the observations showing that the necessary fluctuations can be smaller, such as $\varepsilon_2 = 1.49$ eV/atom [9, 13, 14]. Obviously, the distance between atoms thus should increase.

The third mechanism can be called the "intra-cell" graphitization [15, 16]. In this model the potential energy of a cell has supposedly two local minima. A transition from the higher energy minimum (diamond) to the lower one (graphite) requires only a small fluctuation. The distance between atoms thus should again increase.

The investigations of graphitization are continuing (see, e.g. [17, 18]).

At least two approaches to the description of the graphitization phase change are possible. First, a very complex approach, is in terms of the quantum description of individual transitions of carbon atoms $sp^3 \rightarrow sp^2$. Second one assumes, that the graphitization is phase change, similar melting. In the latter case it is possible to assume that the Lindeman's criterion applies [19-21] and to find the temperature of graphitization. The advantage of this approach is the minimal set of parameters required for the process study. Let us recall Lindeman's formula for the melting temperature:

$$T_m = \frac{x_m^2}{9\hbar^2} Mk\theta^2 r_s^2 \ . \qquad (1)$$

Here $T_m$ is the melting temperature (in our case, the graphitization temperature), $x_m$ is a dimensionless parameter whose value lies in the range of 0.2 to 0.25, $M$ is the atom mass, k is Boltzmann constant, $\theta$ is Debye temperature $(k\theta = \hbar\omega_D)$, $r_s$ is the mean radius of the elementary lattice cell. For the first mechanism $a_d = 1.78 \times 10^{-8}$ cm (mean radiuses of elementary cells of diamond), $a_g = 2 \times 10^{-8}$ cm (mean radiuses of elementary cells of graphite), the activation energy $\varepsilon_1 = 7.6 eV/atom$, $M = 2 \times 10^{-23}$ g (carbon), $\omega_D \approx 2\pi 10^{14} rad/s$, $x_m = 0.21$. For other mechanisms all parameters are the same except for the activation energy $\varepsilon_1 = 1.49\ eV/atom$.

In the first case the transition temperature of phase transition diamond - graphite obtained from the equation (1) is of the order of $10^5$ K. Since at such high temperatures solids turn to plasma, it one can conclude that graphitization as the phase change of the first type is unrealistic.

For the second mechanism, we find from the Lindeman's criterion a similar value for $T_m$. This mechanism is therefore also irrelevant.

A different situation should arise for the "intra-cell" graphitization. Then displacement of a carbon atom is neither $a_g$ nor $a_d$ but is equal to their difference $\delta a = a_g - a_d < 2.2 \times 10^{-9} cm$. In this case from (1) follows $T_m \approx 1360$ K. It is known that such temperatures graphitization rate becomes appreciable [9, 15, 16]. The good agreement of the obtained theoretical estimation and observable value $T_m$ displays, that the using of Lindeman's criterion is justified.

Apparently, we can draw a conclusion that the phase transition "graphitization of diamond" occurs according to the third scenario. However this conclusion demands complementary arguing.

Let's remark that the criterion of Lindeman was offered for the description of a melting. It does not take into account anisotropy of a system and, nevertheless, gives reasonable outcomes. There is no reason to expect that it does not also work for graphitization. As well as the melting, graphitization occurs when the thermal oscillations of atoms reach some characteristic amplitude. It is clear that the overheating of a sample higher then $T_m$ accelerates transitions between minima of a potential energy and can only speed up process of graphitization, as is observed experimentally.

When diamond graphitization is discussed, it is often mentioned that the process occurs in vacuum or in a buffer gas at the normal pressure. The mechanism of "intra-cell" graphitization allows us to estimate the effect of the external pressure on the phase transition temperature. To achieve this, let us consider a diamond plate of some thickness on which an external pressure $P$ is exerted. If the measurements have been carried out at the pressure $P_0$, then increasing the pressure will cause the plate to contract, while decreasing the pressure will cause it to expand. The lattice constant will then change according to

$$L = L_0\left(1 \pm \frac{P}{E}\right). \tag{2}$$

In Equation (2) $E$ is the Young's modulus, the plus sign corresponds to a stretching sample. Since the intra-cell graphitization depends on $\delta a$, we find from (2)

$$\delta a^* = \delta a\left(1 \pm \frac{P}{E}\right). \tag{3}$$

It is still unclear what value of the Young's modulus we should use. This can be an effective elasticity coefficient for the composite system "diamond + graphite", or the Young's modulus of diamond (or one of its modifications, to be precise). From (3) and (1) we find

$$T_m(P) = T_m\left(1 - \frac{P}{E}\right)^2 \approx T_m\left(1 - 2\frac{P}{E}\right), \tag{4}$$

$T_m \approx 1.360 \cdot 10^3 \, \text{K}$ as it found above.

Equation (4) points at an unexpected effect. At the increased pressure the melting temperature is usually increases (known exception is the ice). In (4) this dependence is reversed. This result can be interpreted as follows. The pressure compresses the crystal cell, bringing the higher minimum of energy to the lower one reducing the value of $\delta a = a_g - a_d$ and, according to (1), reducing the transition temperature.

A generalization of several research results is given in [22], providing a plot $P(Torr) = P(T°C)$ based on several data points. The plot shows that as pressure increase, $T_m$ indeed decreases. If one selects several pairs of data points near the normal pressure and considers a function similar to (4), namely

$$T(P) = T_0(1 - \alpha P), \tag{5}$$

then $T_0$ and $a$ can be determined from the plot. It turns out that $T_0 \approx 1400 K \approx T_m$, which is another surprisingly good agreement between our theory and experiments [22]. However, from averaging several data points we find $\alpha \approx 1.5 \times 10^{-7} Pa^{-1}$, which is substantially different from the factor *2/E* in (4). Some possible explanations of this discrepancy are suggested above.

Let us also point out that for low pressures Eq. (4) becomes virtually independent of pressure, which is consistent with the plot from [22].

To conclude, let us point out that that bending of a CVD plate should lead to large mechanical stresses, equivalent to an external compression or stretching in response to externally applied pressure, and to arising of the uphill diffusion, see e.g. [23]. Under these conditions, besides changing of graphitization temperature, re-distribution of impurities should be possible to observe (including cleaning of the CVD diamond surfaces from defects and impurities). Below the $T_m$ this process can be significantly sped up by laser illumination.

## References

1. J. E. Field, *The properties of natural and synthetic diamond*, (Acad. Press, London, 1992).

2. M. A. Prelas, G. Popovici, L.K. Bigelow (eds.), *Handbook of Industrial Diamonds and Diamond Films*, (Marcel Dekker Inc., New York 1997).

3. G. Davies, T. Evans, Proc. R. Soc. London **328**, 413 (1972).

4. De Vita, G. Galli, A.Canning, R. Car, Nature **379**, 523 (1996).

5. V.L.Kuznetsov, I.L.Zilberberg, Yu.V. Butenko, A.L.Chivilin, B. Segall, J. Appl. Phys. **86**, 863 (1999).

6. G. Frieded, G. Ribaud, C.R.Hebd. Seances Acad. Sci, **178**, 1126 (1924).

7. P. Libeau, M. Picon, C.R. Hebd. Seances Acad. Sci, **179**, 1059 (1924).

8. M. Rothschild, C.Arnone, D.J. Ehrlich, J. Vac. Sci. Technol. B, 4, 310 (1986).

9. V.L.Kuznetsov, I.Yu. Malkov, A.L.Chivilin, E.M. Moroz, V.N. Kolomiichuk, Sh.K. Shaichutdinov, Yu.V. Butenko, Carbon, **32**, 873 (1994).

10. V.L. Kuznetsov, A.L. Chivilin, Yu.V. Butenko. I.Yu. Malkov, V.M. Titiv. Chem. Phys. Lett., **222**, 343 (1994).

11. T. Evans, P. James. Proc. R. Soc. London **A 277**, 260 (1964).

12. F.P.Bundy, W.A.Basset, M.S. Weathers, R.J. Hemley, H.K. Mao, A.F. Goncharov. Carbon, **34**, 141 (1996).

13. G. Kern, J. Hafner. Phys. Rev. **B 58**, 13167 (1998).

14. C. Pantea, J. Qian, G. Voronov, T. Zereda. J. Appl. Phys. **91,** 1957 (2002).

15. V.N. Strekalov, D.V. Strekalov. Phys. Rev **B 73**, 115417 (2006).

16. V.N. Strekalov. Appl. Phys. **A 80**, 1061 (2005).

17. Hui Bi, Kai-Chang Kou, Kostya (Ken) Ostrikov, and Jian-Qiang Zhang, J. Appl.Phys. **104,** 033510 (2008).

18. F.J. Ferrer, E. Moreau, D. Vignaud, D. Deresmes, S. Godey, and X. Wallart, J. Appl.Phys. **109,** 054307 (2011).

19. G.A. Lindeman, Phys. Zs., **11**, 609 (1910).

20. J.M. Ziman, *Principles of the theory of solids.* (Cambridge University Press, 1972).

21. C. Kittel, *Introduction to solid state physics.* (Jon Wiley and Sons, Inc, 1956.)

22. V.V. Frunze, A. Yu. Tsutskikh, A.V. Krasil'nikov. Technical Physics Letters **26**, 184 (2000).

23. Ya.E. Geguzin. Sov. Phys. Usp. **29**, 467 (1986).